 \documentclass[sigconf]{acmart}
\AtBeginDocument{%
  }

\copyrightyear{2026}
\acmYear{2026}
\setcopyright{cc}
\setcctype{by}
\acmConference[CIKM '26]{Proceedings of the 35th ACM International Conference on Information and Knowledge Management}{November 07--11, 2026}{Rome, Italy}
\acmBooktitle{Proceedings of the 35th ACM International Conference on Information and Knowledge Management (CIKM '26), November 07--11, 2026, Rome, Italy}
\acmDOI{10.1145/3799682.3839870}
\acmISBN{979-8-4007-2539-5/2026/11}

\usepackage{enumitem}
\usepackage{amsmath}
\usepackage{xspace}
\begin{document}

\title[SPRIG]{SPRIG: Semantic-ID-enhanced Paths for Knowledge Graph-based Generative Recommendation} 

\author{Justin Hangöbl}
\email{hangoebl.j@gmail.com}
\orcid{0009-0004-8304-3985}
\affiliation{%
  \institution{Institute of Computational Perception, Johannes Kepler University} 
  \city{Linz}
  \country{Austria}
}

\author{Marta Moscati}
\email{marta.moscati@jku.at}
\orcid{0000-0002-5541-4919}
\affiliation{\institution{Institute of Computational Perception, Johannes Kepler University}\city{Linz}\country{Austria}}
\affiliation{\institution{Albatross AI}\city{Linz}\country{Austria}}

\author{Alessandro B. Melchiorre}
\email{alessandro.b.mel@gmail.com}
\orcid{0000-0003-1643-1166}
\affiliation{%
  \institution{Criteo AI Lab}
  \city{Paris}
  \country{France}
}

\author{Shah Nawaz}
\email{shah.nawaz@jku.at}
\orcid{0000-0002-7715-4409}
\affiliation{%
  \institution{Institute of Computational Perception, Johannes Kepler University}
  \city{Linz}
  \country{Austria}}

\author{Markus Schedl}
\email{markus.schedl@jku.at}
\orcid{0000-0003-1706-3406}
\affiliation{%
  \institution{Institute of Computational Perception, Johannes Kepler University}
  \city{Linz}
  \country{Austria}}
  \affiliation{
  \institution{Human-centered AI Group, AI Lab, Linz Institute of Technology}
  \city{Linz}
  \country{Austria}}

\newcommand{\ncold}{\ensuremath{n_\text{cold}}}

\newcommand{\repourl}{\url{https://github.com/justinhangoebl/semantic-id-knowledge-graph-recommender}\xspace}
\newcommand{\ie}{i.\,e., }
\newcommand{\wrt}{w.\,r.\,t.\, }
\newcommand{\eg}{e.\,g., }
\newcommand{\cf}{cf.\, }
\newcommand{\st}{s.\,t.\, }
\newcommand{\lfmo}{Onion\xspace}
\newcommand{\movielens}{ML-1M\xspace}


\newif\ifworkinprogress
\workinprogressfalse

\ifworkinprogress
	\newcommand{\ms}[1]{\textcolor{blue}{{[Markus] #1}}}
	\newcommand{\mm}[1]{\textcolor{olive}{{[Marta] #1}}}
	\newcommand{\jh}[1]{\textcolor{green}{{[Justin] #1}}}
	\newcommand{\abm}[1]{\textcolor{orange}{{[Alessandro] #1}}}
	\newcommand{\abm}[1]{\textcolor{red}{{[Shah] #1}}}
\else
    \newcommand{\ms}[1]{}
    \newcommand{\mm}[1]{}
    \newcommand{\jh}[1]{}
    \newcommand{\abm}[1]{}
\fi

\renewcommand{\shortauthors}{Justin Hangöbl, Marta Moscati, Alessandro B. Melchiorre, Shah Nawaz, and Markus Schedl}

\newcommand{\onion}{\text{Onion}\xspace}
\newcommand{\amazon}{\text{Amazon}\xspace}
\newcommand{\ease}{\text{EASE}\xspace}
\newcommand{\multvae}{\text{MultVAE}\xspace}
\newcommand{\multdae}{\text{MultDAE}\xspace}
\newcommand{\als}{\text{ALS}\xspace}
\newcommand{\iknn}{\text{Item-}$k$\text{NN}\xspace}
\newcommand{\bert}{\text{BERT4Rec}\xspace}
\newcommand{\sasrec}{\text{SASRec}\xspace}
\newcommand{\rand}{\text{Rand}\xspace}
\newcommand{\pop}{\text{Pop}\xspace}
\newcommand{\ul}[1]{\underline{#1}}


\begin{abstract}
Recommender systems leveraging generative models often generate item identifiers directly, rather than ranking catalog items by a recommendation score.
Recent work extends beyond pure sequential interaction signals by incorporating item content and structured relationships among items, with two distinct directions emerging.
\emph{Semantic IDs} (SIDs) enrich item representations by replacing opaque, randomly initialized embeddings with hierarchically quantized discrete codes derived from item content. \emph{Knowledge-graph} (KG) path reasoning instead generates entity-relation paths that ground recommendations in structured relationships between items, attributes, and external entities, thereby enriching the relational context. These two lines have complementary limitations: SID-based models lack relational grounding, while KG-based generative recommenders still represent items as arbitrary, opaque tokens tied to large embedding tables, limiting parameter sharing and generalization. We propose \textbf{SPRIG}, a generative recommender that integrates content-derived SIDs into KG path reasoning. SPRIG is trained on information-rich KG paths that terminate in items represented as discrete, content-derived tokens, combining the advantages of both approaches. We evaluate SPRIG on movie and music recommendation datasets against baselines spanning sequential language models, KG-augmented methods, and SID-based approaches. Our results show that SPRIG achieves competitive performance over prior generative models while using fewer parameters and a lower compute cost.
Code: \repourl

\end{abstract}

\ccsdesc[500]{Information systems~Collaborative filtering}
\ccsdesc[300]{Information systems~Personalization}
\ccsdesc[500]{Information systems~Recommender systems}
\ccsdesc[500]{Information systems~Multimedia and multimodal retrieval}
\keywords{Generative Recommendation,
Path reasoning,
Path language modelling,
Knowledge graph recommendation,
Semantic identifiers,
Hierarchical quantization,
Residual quantization
}

\maketitle

\begin{figure}[ht]
    \centering
    \includegraphics[width=\linewidth]{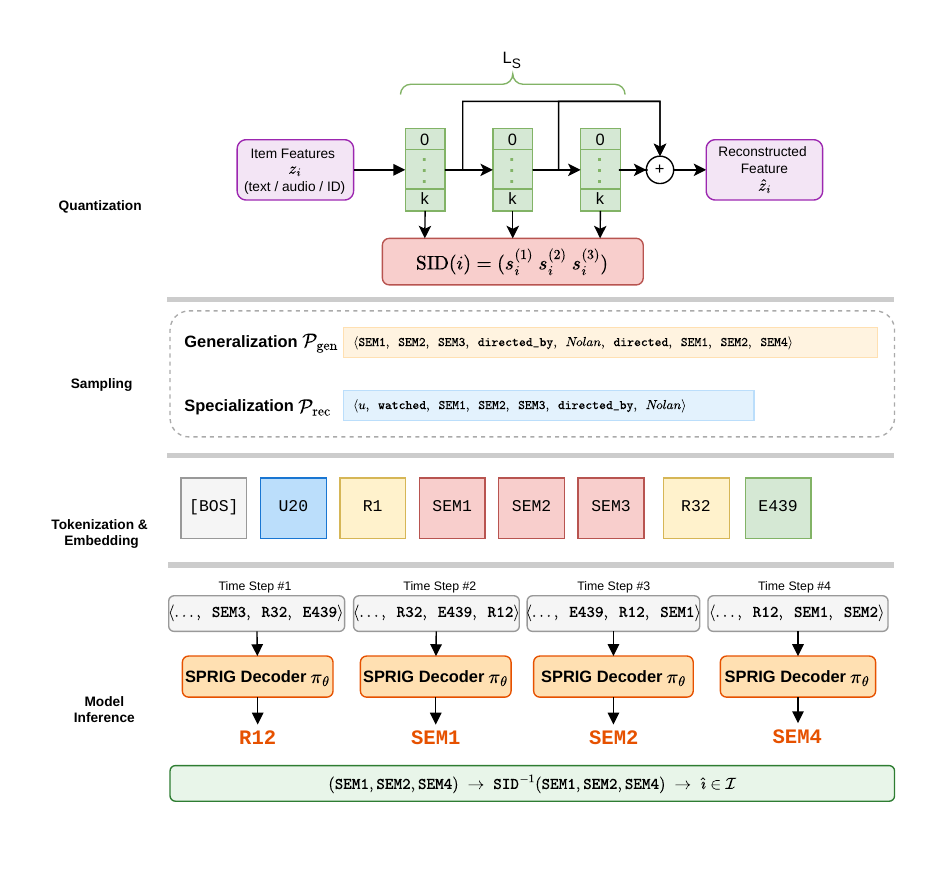}
    \caption{SPRIG framework. Item features are encoded into hierarchical Semantic IDs; SID-augmented KG paths (generic for pretraining, user-to-item for fine-tuning) are tokenized and fed to a causal Transformer decoder that autoregressively generates paths. Section~\ref{sec:method} follows a concrete movie example (\emph{Inception}, \emph{Interstellar}) through all four stages.}
    \label{fig:architecture}
    \Description{A four-stage pipeline diagram, read top to bottom, with stages Quantization, Sampling, Tokenization and Embedding, and Model Inference. In the Quantization stage, an item feature vector derived from text, audio, or ID passes through three codebooks in series, each indexing entries from zero to k. The same three selected codes, form the item's Semantic ID: an ordered triple of one code per level. In the Sampling stage, two kinds of knowledge-graph paths are drawn. A generalization path interleaves entities, relations, and Semantic ID codes across a long walk; a shorter specialization path starts at a user node and ends at a target item entity. In the Tokenization and Embedding stage, one such path is laid out as a flat token sequence, a user token, alternating relation tokens, Semantic ID code tokens, and an entity token. In the Model Inference stage, four successive time steps are shown. At each step tokens are fed to the SPRIG decoder, which emits one token; across the four steps it emits a relation token followed by the three Semantic ID codes of the predicted item. An inverse Semantic ID lookup maps that code triple back to a single item in the catalog.}
\end{figure}

\section{Introduction and Related Work} \label{sec:intro}
    
    Recommender systems have recently shifted from score-based retrieval towards \emph{generative} formulations, where a causal Transformer model trained on chronologically ordered interaction sequences autoregressively predicts the next item directly from a user's history~\cite{geng2022p5,rajput2023tiger,zhai2024hstu}. 
    While these advances brought substantial gains, two limitations persist: the vocabulary size required for items grows linearly with catalog size, and the sequential (collaborative) signal alone is not rich enough. Complementary directions have emerged to inject richer contextual information into generative recommendation.
    The first approach enriches the item representation by replacing the arbitrary, opaque, randomly-initialized item IDs with semantically rich discrete codes, known as Semantic IDs (SIDs)~\cite{rajput2023tiger,hou2023vqrec, tan2024idgenrec, clark2025grid}. Because each item is encoded with codes from a small shared codebook, SIDs also substantially reduce the vocabulary size.
    The second enriches the relational context by grounding recommendation in \emph{knowledge graphs} (KGs), which encode structured relationships between items, attributes, and external entities~\cite{KGGLM, xian2019pgpr}.
    
    \textit{Semantic IDs for Generative Recommendation.} 
        SIDs were popularized by TIGER~\cite{rajput2023tiger}, which represents each item as a short sequence of content-derived codes from an RQ-VAE over pretrained embeddings~\cite{lee2022rqvae,rajput2023tiger}. Sharing code prefixes across semantically similar items enables parameter sharing and cold-start generalization~\cite{singh2024betterids}, and later work adds collaborative signals~\cite{wang2024letter,zheng2024lcrec} or end-to-end tokenization~\cite{liu2025etegrec}. SPRIG adopts the TIGER formulation with the refined quantization of Generative Recommendation with Semantic IDs (GRID)~\cite{clark2025grid}, but unlike these methods grounds recommendation in a structured relational context.
    
    \textit{Knowledge-Graph-Augmented Generative Recommendation.} 
        A parallel line frames recommendation as \emph{explicit path reasoning} over a KG, returning an item together with a human-interpretable multi-hop path. Early reinforcement-learning (RL)-based agents navigate from user to item~\cite{xian2019pgpr,balloccu2023eqpgr,balloccu2022rlpath}; the paradigm was later recast as \emph{autoregressive language modeling over KG paths}, with constrained decoding guaranteeing valid edges~\cite{geng2022plm,balloccu2023pearlm}. 
        KGGLM~\cite{KGGLM} is the state of the art in this line, pretraining a generative model on generic KG paths and fine-tuning for path-reasoning recommendation. However, all of these represent items as opaque, arbitrary IDs, preventing parameter sharing and growing the vocabulary linearly with the catalog size.
    
        In this context, we propose \textbf{SPRIG}, a generative recommender framework that unifies content-derived SIDs with knowledge-graph path reasoning in a single autoregressive model. SPRIG ingests hierarchical SIDs pre-computed from item content features via a residual quantizer, replacing the opaque atomic item tokens of prior KG-based recommenders with semantically informed, parameter-sharing codes. Afterwards, it leverages the rich contextual structure of a KG to pretrain a causal Transformer decoder on generic entity-relation paths, before fine-tuning on user-to-item recommendation paths whose last tokens consist of SID sequences. The path generator builds on KGGLM~\cite{KGGLM}, which we also adopt as our primary KG baseline. This design simultaneously inherits the expressiveness of KG path reasoning~\cite{xian2019pgpr,geng2022plm,balloccu2023pearlm,KGGLM} and the parameter efficiency and cross-item generalization of SIDs~\cite{rajput2023tiger,wang2024letter}, while decoupling the output vocabulary from the catalog size. To the best of our knowledge, SPRIG is the first model to integrate SIDs with KG path reasoning for generative recommendation. Within this context, we investigate whether KG-based generative recommenders benefit from representing item entities using semantically grounded identifiers (\textbf{RQ1}), and whether the resulting shared code space reduces the output vocabulary without sacrificing recommendation quality (\textbf{RQ2}). To answer these questions we make the following contributions:
        \begin{itemize}[
              topsep=2pt,       
              partopsep=0pt,    
              itemsep=2pt,      
              leftmargin=10pt,
              parsep=0pt        
            ]
            \item We propose SPRIG, the first generative recommender that unifies SIDs and KG path reasoning in a single autoregressive framework.
            \item We show in two recommendation domains (movies on \movielens and music on \lfmo) that SPRIG achieves competitive performance against sequential language models, KG-augmented methods, and SID-based baselines.
        \end{itemize}
    
\section{Methodology} \label{sec:method}

    SPRIG extends the path-language-modeling paradigm of KGGLM~\cite{KGGLM} by incorporating content-aware Semantic IDs inspired by TIGER~\cite{rajput2023tiger}, replacing atomic item identifiers with structured code sequences within a single shared vocabulary.

    SPRIG follows a three-stage design, illustrated in Figure~\ref{fig:architecture}. First, item content features are quantized into hierarchical SIDs (\eg \emph{Inception} maps to $(\texttt{SEM1},\texttt{SEM2},\texttt{SEM3})$ and the similar \emph{Interstellar} to $(\texttt{SEM1},\texttt{SEM2},\texttt{SEM4})$, sharing the prefix $(\texttt{SEM1},\texttt{SEM2})$, while the unrelated \emph{Barbie} maps to $(\texttt{SEM5},\texttt{SEM6},\texttt{SEM7})$). These codes are then used to represent item entities within KG paths by expanding each item occurrence into its corresponding SID sequence (so $\langle u,\,\texttt{watched},\, \textit{Inception}\rangle$ becomes $\langle u,\,\texttt{watched},\,\texttt{SEM1},\, \texttt{SEM2},\,\texttt{SEM3}\rangle$), yielding a unified vocabulary of users, relations, attributes, and SID tokens.

    A causal Transformer decoder is subsequently trained in two stages: \textbf{pretraining} on generic KG paths to learn relational structure without any collaborative signal (\eg $\langle$\,$\texttt{SEM1},\texttt{SEM2},\texttt{SEM3}$, \\ $\texttt{directed\_by}$,\textit{Nolan}, $\texttt{directed}$, $\texttt{SEM1},\texttt{SEM2},\texttt{SEM4}$\,$\rangle$), followed by \textbf{finetuning} on user-to-item paths to specialize the model for recommendation (\eg $\langle$\,$u$, $\texttt{watched}$, $\texttt{SEM1},\texttt{SEM2},\texttt{SEM3}$, $\texttt{directed\_by}$, \textit{Nolan}, $\texttt{directed}$ , $\texttt{SEM1},\texttt{SEM2},\texttt{SEM4}$ \, $\rangle$). At inference, given a user $u$ and the \texttt{watched} relation, the model autoregressively emits a SID sequence, \eg $\texttt{SEM1},\texttt{SEM2},\texttt{SEM4})$, which an inverse SID lookup maps back to a catalog item (\emph{Interstellar}).
    
    \subsection{Preliminaries}

            Let $\mathcal{U}$, $\mathcal{I}$, and $\mathcal{A}$ denote the sets of users, items, and attribute entities, respectively. SPRIG operates on a heterogeneous knowledge graph $\mathcal{G}=(\mathcal{E},\mathcal{R},\mathcal{T})$ with entities $\mathcal{E}=\mathcal{U}\cup\mathcal{I}\cup\mathcal{A}$, relation types $\mathcal{R}$, and triplets $\mathcal{T}\subseteq\mathcal{E}\times\mathcal{R}\times\mathcal{E}$. User-item interactions are encoded as triplets $(u,{\mathrm{UI}},i)\in\mathcal{T}$.
            
            A \emph{KG path} of length $L$ is a sequence $p = e_0\,r_1\,e_1\,\dots\,r_L\,e_L$ with $(e_{\ell-1},r_\ell,e_\ell)\in\mathcal{T}$ for all $\ell$. SPRIG is trained on two corpora of such paths, sampled differently for pretraining and finetuning; since the sampling strategy is part of our approach, we define both in Section~\ref{sec:paths}.

            Following TIGER and the broader SID literature~\cite{rajput2023tiger,clark2025grid}, a hierarchical quantizer with codebook size $K$ represents an item $i$ as a sequence $\mathrm{SID}(i) = \big(s^{(1)}_i,\,\dots,\,s^{(L_s)}_i\big)$ of $L_s$ discrete codes $s^{(\ell)}_i \in \{1,\dots,K\}$ derived from a content embedding $z_i$, where each successive code refines the residual of the previous one. Because semantically similar items thus share longer code prefixes, this representation enables parameter sharing across related items~\cite{rajput2023tiger}. Opaque identifiers cannot express this similarity: in KGGLM, \texttt{I123} and \texttt{I124} are adjacent tokens denoting unrelated items, and nothing in an identifier reflects that \emph{Inception} and \emph{Interstellar} are alike. SIDs instead express similarity level by level: \emph{Inception} $(\texttt{SEM1},\texttt{SEM2},\texttt{SEM3})$ and \emph{Interstellar} $(\texttt{SEM1},\texttt{SEM2},\texttt{SEM4})$ agree on the coarse codes and differ only in the last, while \emph{Barbie} $(\texttt{SEM5},\texttt{SEM6},\texttt{SEM7})$ diverges at the first, so SPRIG can share parameters across related items. SPRIG is agnostic to the quantizer and compatible with any scheme producing $\mathrm{SID}(i)$, \eg RVQ~\cite{zeghidour2021soundstream}, residual K-means~\cite{clark2025grid}, RQ-VAE~\cite{lee2022rqvae,rajput2023tiger}.            
        
    \subsection{SID-augmented Knowledge Graph Paths} \label{sec:paths}

        SPRIG draws two path corpora from the same graph, mirroring the pretrain-then-finetune paradigm of KGGLM~\cite{KGGLM}. For \emph{pretraining}, we sample paths that start from an arbitrary item or attribute entity ($e_0\in\mathcal{I}\cup\mathcal{A}$), traverse only KG relations (user-item edges excluded), and have unconstrained endpoints, yielding a corpus $\mathcal{P}_{\mathrm{gen}}$ that captures relational structure without any collaborative signal. For \emph{finetuning}, we sample paths that start at a user ($e_0=u\in\mathcal{U}$), traverse the full graph including user-item edges, and terminate at a target item ($e_L=i\in\mathcal{I}$), this yields the corpus $\mathcal{P}_{\mathrm{rec}}$.

        The central design choice of SPRIG is to inject the SIDs directly into the KG path vocabulary. Given a raw KG path $p = \langle e_0, r_1, e_1,\dots,r_L,e_L\rangle$, we define its \emph{SID-augmented rewriting} $\tilde{p}$ by replacing every item entity $e_\ell\in\mathcal{I}$ with its SID sequence:
        \begin{equation}
            \tilde{p} = \langle \tilde{e}_0\,r_1\,\tilde{e}_1\,\dots\,r_L\,\tilde{e}_L \rangle
            \qquad
            \tilde{e}_\ell =
            \begin{cases}
                \mathrm{SID}(e_\ell) & \text{if } e_\ell\in\mathcal{I},\\
                e_\ell & \text{otherwise}.
            \end{cases}
            \label{eq:rewrite}
        \end{equation}
        Since every item occurrence is expressed as a tuple of SID tokens, a graph path of $L$ hops is represented as a longer token sequence $\tilde{p}=\langle t_1,\dots,t_T\rangle$ with $T \geq L$, on which SPRIG operates rather than on the graph path directly.
        
        \textit{Unified vocabulary.} %
            All paths, generic and recommendation alike, share a single vocabulary given by $\mathcal{V} = \mathcal{V}_{\mathrm{sid}} \cup \mathcal{V}_{\mathrm{ent}} \cup \mathcal{V}_{\mathrm{rel}} \cup \mathcal{V}_{\mathrm{special}}$,
            where $\mathcal{V}_{\mathrm{sid}}$ is the set of $L_s \cdot K$ SID tokens  ($K$ codes per level across $L_s$ levels), $\mathcal{V}_{\mathrm{ent}}=\mathcal{U}\cup\mathcal{A}$ the atomic non-item entities, $\mathcal{V}_{\mathrm{rel}}=\mathcal{R}$ the relation types, and $\mathcal{V}_{\mathrm{special}}$ the special tokens (\texttt{<bos>}, \texttt{<eos>}, \texttt{<pad>}). The total vocabulary size is
            $|\mathcal{V}| = L_s \cdot K + |\mathcal{U}| + |\mathcal{A}| + |\mathcal{R}| + |\mathcal{V}_{\mathrm{special}}|,$
            in which the item-dependent component is the \emph{constant} $L_s \cdot K$ rather than growing as $|\mathcal{I}|$. This decouples the vocabulary size from the catalog size, allowing SPRIG to scale to large catalogs while retaining the expressiveness of KG path reasoning, in contrast to PGPR-style models~\cite{xian2019pgpr,balloccu2023pearlm,KGGLM} whose vocabulary grows linearly with~$|\mathcal{I}|$.

    \subsection{Decoder Model Training} \label{sec:training}
        Given the SID-augmented vocabulary $\mathcal{V}$, SPRIG models a tokenized path $\tilde{p}=\langle t_1,\dots,t_T\rangle$ with a causal Transformer decoder $\pi_\theta$ that factorizes the sequence autoregressively, $\pi_\theta(\tilde{p})=\prod_{j=1}^{T}\pi_\theta(t_j\mid t_{<j})$, as in KGGLM~\cite{KGGLM} and PEARLM~\cite{balloccu2023pearlm}. Over a path corpus $\mathcal{P}$, the parameters are trained to minimize the negative log-likelihood.
        \begin{equation}
            \mathcal{L}(\theta;\mathcal{P})
            = -\frac{1}{|\mathcal{P}|}\sum_{\tilde{p}\in\mathcal{P}}\;\sum_{j=1}^{T}
              \log \pi_\theta\!\left(t_j \mid t_{<j}\right).
            \label{eq:loss}
        \end{equation}
        Because the same objective is applied to every token, whether it denotes a user, relation, attribute, or SID code, SPRIG learns a single joint distribution over the relational vocabulary and the hierarchical item codes. Training proceeds in two stages over the same graph, differing only in the corpus $\mathcal{P}$ from which paths are sampled.
    
        \textit{Pretraining.}
            We first sample a generic corpus $\mathcal{P}_{\mathrm{gen}}$ of paths that start from an arbitrary item or attribute entity ($e_0\in\mathcal{I}\cup\mathcal{A}$), traverse only KG relations (user-item edges excluded), and have unconstrained endpoints, and minimize $\mathcal{L}(\theta;\mathcal{P}_{\mathrm{gen}})$. Mirroring KGGLM's generic pretraining~\cite{KGGLM}, this stage uses no user-item interaction as supervision; it lets the model absorb structural regularities of the graph, such as frequent relation chains and co-occurrence between attribute entities and SID prefixes.
    
        \textit{Finetuning.}
            We then sample a recommendation corpus $\mathcal{P}_{\mathrm{rec}}$ of paths that start at a user ($e_0=u\in\mathcal{U}$), traverse the full graph including user-item edges, and terminate at a target item ($e_L=i\in\mathcal{I}$). Continuing from the pretrained parameters, we minimize $\mathcal{L}(\theta;\mathcal{P}_{\mathrm{rec}})$, so the model learns to place weight on paths that both (i) respect the relational structure of the KG and (ii) terminate in a valid SID sequence identifying an item consumed by the user.

\section{Experiments}
    In this section, we describe the setup of our experiments, which aim to uncover the effects of supplementing KG path reasoning with SIDs on recommendation accuracy and model size. 
    
    \textit{Datasets. }%
        \begin{table}[t]
            \centering
            \caption{Dataset Characteristics of \movielens and \lfmo}
            \label{tab:dataset_characteristics}
            \resizebox{\columnwidth}{!}{\begin{tabular}{lrr|lrr}
                \toprule
                \multicolumn{3}{c|}{Collaborative Components} & \multicolumn{3}{|c}{Knowledge Components} \\
                \midrule
                Characteristic & \movielens & \lfmo & Characteristic & \movielens & \lfmo \\
                \midrule
                Users                 &  6,040      &  12,784     & Entities (Types)        & 13,804 (12)  & 26,701 (6)  \\
                Items                 &  2,984      & 18,415      & Relations (Types)       & 193,089 (11) & 172,232 (6) \\
                Interactions          & 932,295     & 1,375,341   & Avg.\ Degree (All)      & 28.07        & 12.90        \\
                Density               & 0.05        & 0.01        & Avg.\ Degree (Items)    & 64.86        & 9.35        \\
                \bottomrule
            \end{tabular}}
        \end{table}
        Following KGGLM~\cite{KGGLM}, we evaluate on two publicly available datasets spanning distinct domains: MovieLens-1M (\movielens)~\cite{zhao2022recbole} for movie recommendation and LFM-2B~\cite{schedl2022lfm} for music track recommendation. We adopt the KG-enriched version of \movielens available via RecBole~\cite{zhao2022recbole}. We complement LFM-2B with the music content features from~\cite{moscati2022onion} and use artist, genre, album, and user-generated tags from the same dataset, to construct the KG. Given the scale of LFM-2B, we restrict to interactions in March--May 2018 and apply iterative $k$-core filtering ($k=20$) until the user and track sets stabilize, we call this combined dataset \lfmo. Detailed statistics for both datasets after pre-processing are reported in Table~\ref{tab:dataset_characteristics}.

        SIDs are generated offline with residual K-means~\cite{clark2025grid} ($L_s{=}3$ levels, $K{=}256$ codes) on T5 / musicnn embeddings for \movielens / \lfmo, which performs best in our ablations (Sec.~\ref{sec:ablations}), in agreement with GRID~\cite{clark2025grid}. This yields an item-token vocabulary of $L_s \cdot K = 3 \cdot 256 = 768$ shared codes, versus the $|\mathcal{I}|{=}2{,}984$ (\movielens) and $18{,}415$ (\lfmo) item embeddings required by KGGLM.

        \textit{Implementation Details.}
            All experiments use an NVIDIA A40 GPU. Each model is run under its own best configuration from the original publications~\cite{KGGLM, rajput2023tiger, kang2018sasrec, OpenP5} and verified on our splits, with identical dataset preprocessing and chronological splits across models. The DistilGPT-2 backbone (6 layers, 12 heads, $d{=}768$) is used off-the-shelf and shared by SPRIG and KGGLM, so it is not tuned in SPRIG's favor; SPRIG only lowers the learning rate and shortens the path length, as the longer $L_s$-token SIDs require. We pretrain on 500 5-hop KG random walks per entity for 3 epochs, then fine-tune on collaborative 3-hop paths (200 per user).

        \textit{Task and Protocol.}
            SPRIG performs top-k recommendation by path reasoning: conditioned on a user ($\langle[BOS], U_{id}, R_{UI} \rangle$), it generates user-to-item paths via diverse beam search and ranks the terminal items. We use a per-user chronological 60/20/20 split and evaluate under full ranking over the catalog against each user's held-out interactions.
            We use diverse beam search with $60$ beams in $5$ groups, diversity penalty $0.4$, and length penalty $0$, deterministically generating $3$-hop paths and returning up to $50$ candidate paths per user. Terminal items are ranked by their path log-likelihood; when several beams reach the same item we keep its highest-scoring path. Because SPRIG and KGGLM share this decoding configuration, their comparison is controlled, whereas SASRec ranks the full catalog directly and therefore serves as a reference point rather than a head-to-head generative baseline.
        
        \textit{Evaluation Metrics.} %
            We evaluate ranking quality with three standard metrics at $K=10$. We assess statistical significance with a two-sided paired bootstrap test over per-user metric scores, comparing SPRIG (gen + spec) against KGGLM (gen + spec).

        \textit{Baselines.} 
        We compare SPRIG against baselines spanning sequential, generative, and knowledge-aware recommendation paradigms. SASRec~\cite{kang2018sasrec} is a strong sequential recommendation baseline based on self-attentive sequence modeling. Unlike the generative methods below, it directly ranks items from the full catalog rather than generating recommendation paths or identifiers. OpenP5~\cite{OpenP5} is a generative recommender that represents items with atomic identifiers; we report both its sequential and collaborative indexing variants. TIGER~\cite{rajput2023tiger} represents items with semantically grounded codes derived from item content, isolating the effect of SIDs in the identifier space. KGGLM~\cite{KGGLM} is the direct predecessor of SPRIG and our strongest knowledge-aware baseline: it incorporates KG structure but retains opaque item tokens, so the gap between KGGLM and SPRIG isolates the contribution of our SID scheme rather than the KG itself. For both KGGLM and SPRIG we report a generalization-only variant (\emph{gen-only}), trained solely on the generative objective, and a specialized variant (\emph{gen + spec}), further fine-tuned for recommendation.
    
\section{Results} \label{sec:experiments}

    Table \ref{tab:results} reports the performance of SPRIG and all baselines on \movielens and \lfmo. SPRIG (gen + spec) performs best among directly comparable generative models. We analyze these results to answer the RQs from Section \ref{sec:intro}.
    
    \begin{table}[t]
        \centering
        \caption{Models performance on \movielens and \lfmo; all metrics $@10$. Bold: best, \ul{underlined}: second best. $^\ast$ marks SPRIG (gen + spec) results significantly better than KGGLM (gen + spec) (paired bootstrap, $p<0.05$).}
        \label{tab:results}
        \resizebox{\columnwidth}{!}{
            \begin{tabular}{lccc|ccc}
                \toprule
                 & \multicolumn{3}{c}{\movielens} & \multicolumn{3}{c}{\lfmo} \\
                \cmidrule(r{0.25em}){2-4}\cmidrule(l{0.25em}){5-7}
                Model                           & nDCG   & MRR       & HR & nDCG & MRR    & HR \\
                \midrule
                TopPop                          & 0.1086        & 0.1251        & 0.4396          & 0.0195          & 0.0487        & 0.1361            \\  
                SASRec \cite{kang2018sasrec}    &\textbf{0.1261}& 0.0912        & 0.2416          &\textbf{0.1892}  &\ul{0.1514}    &\textbf{0.3072}    \\
                OpenP5 (seq.) \cite{OpenP5}     & 0.0243        & 0.0158        & 0.0542          & 0.0029          & 0.0022        & 0.0050            \\ 
                OpenP5 (coll.) \cite{OpenP5}    & 0.0192        & 0.0130        & 0.0397          & 0.0185          & 0.0121        & 0.0401            \\
                TIGER (GRID Variant) \cite{rajput2023tiger, clark2025grid}      & 0.0565          & 0.1324        & 0.2413          & 0.0252          & 0.0182        & 0.0480            \\
                KGGLM (gen-only) \cite{KGGLM}   & 0.0281        & 0.0709        & 0.1930          & 0.0134          & 0.0301        & 0.0770            \\
                KGGLM (gen + spec) \cite{KGGLM} & 0.0864        &\ul{0.1919}    & \ul{0.4407}     & 0.0421          & 0.0911        & 0.2223            \\
                \midrule
                SPRIG (gen-only)                & 0.0375        & 0.0940        & 0.2879          & 0.0140           & 0.0433        & 0.0865            \\
                SPRIG (gen + spec)              &\ul{0.1241}$^\ast$&\textbf{0.2874}$^\ast$&\textbf{0.5632}$^\ast$&\ul{0.0669}$^\ast$  &\textbf{0.1738}$^\ast$&\ul{0.2653}$^\ast$       \\
                \bottomrule
            \end{tabular}
        }
    \end{table}
    
        We first address \textbf{RQ1}: whether replacing the opaque item tokens used in KG-based generative recommenders with semantically-grounded IDs improves recommendation quality. On \movielens the fully specialized SPRIG (gen + spec) improves nDCG@10 by roughly $44\%$ over its directly comparable baseline KGGLM (gen + spec). The same trend holds on the larger and sparser \lfmo, where SPRIG raises nDCG@10 by $58\%$. Since KGGLM and SPRIG share the same backbone and knowledge-graph signal, this gain stems from the SID-based item representation and its associated decoding rather than from the knowledge graph, answering RQ1 affirmatively. These improvements over KGGLM (gen + spec) are statistically significant across all metrics on both datasets (Table~\ref{tab:results}).

        Across representation paradigms, the generative baselines OpenP5 and TIGER are substantially outperformed by the knowledge-aware models, confirming that relational grounding adds signal that SID-based sequential models alone miss. SASRec attains the highest nDCG@10 on both datasets and the highest HR@10 on \lfmo, but it ranks the full catalog directly while the generative models decode over beam-search candidates, so the comparison is indicative rather than exact; its margin on \lfmo further reflects the sequentiality of music listening that the sampled setup discards. Among directly comparable generative models, the gap between SPRIG and KGGLM shows that relational grounding and SIDs are complementary and the combination of them is strongest.

        Even without collaborative fine-tuning, SPRIG (gen-only) surpasses KGGLM (gen-only) on every metric on both datasets (\eg nDCG@10 improves from $0.0281$ to $0.0375$ on \movielens). We attribute this to the parameter sharing induced by SIDs, which lets rarely seen items inherit signal from their prefix-level neighbors; the advantage widens further after specialization.

        We now turn to \textbf{RQ2}: whether representing items in a smaller, shared latent code space allows a generative recommender to reduce its output vocabulary while preserving accuracy. KGGLM allocates one embedding per item, so its item-dependent vocabulary grows linearly with the catalog ($|\mathcal{I}|=2{,}984$ for \movielens and $18{,}415$ for \lfmo). SPRIG instead represents every item with $L_s=3$ codes drawn from $K=256$ shared entries, bounding the item-token vocabulary by the factor $L_s \cdot K=768$ regardless of catalog size. On \movielens this shrinks the item-token budget from $|\mathcal{I}|=2{,}984$ to $L_s \cdot K = 768$ (a $\sim4\times$ reduction), and the saving widens to $\sim24\times$ on the larger \lfmo catalog ($18{,}415$ to $768$).

        This compression does not cost capacity. On \movielens, SPRIG is comparable in size to KGGLM ($56.1$M vs.\ $57.8$M parameters) yet uses $11\%$ fewer FLOPs per step, while reaching the higher accuracy in Table~\ref{tab:results}. This gap grows with the catalog. On \lfmo, where KGGLM's per-item embeddings dominate its parameter count, SPRIG uses $19\%$ fewer parameters and nearly half the FLOPs. Decoupling the vocabulary from the catalog thus improves both accuracy and efficiency, and does so most where the catalog is largest.

        \paragraph{Ablation: Robustness to Quantization and Item Representation}\label{sec:ablations}
            To test the robustness of SPRIG to the quantization approach, we evaluate three residual quantizers for semantic-ID construction: RVQ~\cite{zeghidour2021soundstream}, RQ-VAE~\cite{lee2022rqvae, rajput2023tiger}, and residual K-means~\cite{clark2025grid}. Consistent with GRID~\cite{clark2025grid}, residual K-means attains the best accuracy, but the gap over RVQ and RQ-VAE is small ($\sim$1\%), indicating that SPRIG is largely insensitive to the choice of quantizer. We then test robustness to the item representation by comparing two audio feature extractors on \lfmo. Here the behavior differs: musicnn embeddings~\cite{pons2019musicnn} consistently outperform the MFCC bag-of-words~\cite{hossan2010mfcc} representation across all quantizers, showing that SPRIG is instead sensitive to the quality of the underlying item features.
    
\section{Conclusions and Future Work}

    We introduced SPRIG, the first generative recommender to unify content-derived SIDs with KG path reasoning in a single autoregressive framework. By rewriting every item entity in the KG as a short sequence of hierarchical tokens and training a causal Transformer decoder over the resulting paths, SPRIG combines the relational reasoning of KG-based recommenders with the parameter efficiency and cross-item generalization of SIDs. Across two domains --- movies on \movielens and music on \lfmo --- SPRIG outperforms prior generative recommenders spanning atomic-ID (OpenP5), SID-based (TIGER), and KG-augmented (KGGLM) families, surpassing its direct predecessor KGGLM on every metric, while remaining competitive with a strong sequential baseline (SASRec) at a lower parameter count and decoupling the output vocabulary from the catalog size.

    Several directions remain open. First, the use of SIDs offers as natural and promising application the use of SPRIG for item cold-start recommendation: a new item can be reached through the SID prefix it inherits from content-similar seen items, rather than requiring a dedicated embedding~\cite{rajput2023tiger, singh2024betterids}. A faithful evaluation, however, requires cold-start–specific splits and a path-sampling strategy for items absent from the user–item training edges, which we leave to future work. Second, while KG path reasoning is inherently explanatory, SPRIG's paths terminate in SID codes that are less human-readable than named item entities; recovering faithful, interpretable justifications(\eg by aligning the hierarchical code levels with the attribute entities that generated them) is a promising but non-trivial direction rather than an existing feature of the model. In addition, our finding that the optimal quantizer differs from that reported for purely sequential SID models~\cite{clark2025grid} motivates a deeper study of how tokenizer choice interacts with relational path modeling. We leave these research directions for future work.

\begin{acks}
This research was funded in whole or in part by the Austrian Science Fund (FWF): \href{https://doi.org/10.55776/COE12}{10.55776/COE12}, \href{https://doi.org/10.55776/DFH23}{10.55776/DFH23}, \href{https://doi.org/10.55776/P36413}{10.55776/P36413}.
\end{acks}

\section*{Generative AI Disclosure}

In accordance with the ACM Policy on Authorship, the authors disclose the use of a large language model (Anthropic Claude, Opus 4 and Opus 5 model families) during the preparation of this work. The tool was used for (i) linguistic refinement and stylistic harmonization of author-written text, and (ii) coding assistance, including editor auto-complete, boilerplate components of standard, well-established routines (e.g., file serialization) according to author-provided specifications. All conceptual design, experimental protocols, analysis, and reported results are the authors' own. All AI-assisted text and code was reviewed, tested, and verified by the authors, who take full responsibility for the content of this work.

\bibliographystyle{ACM-Reference-Format}
\balance
\bibliography{bibliography}

\end{document}